\documentclass[prb,twocolumn,amsmath,amssymb,showpacs,times]{revtex4}
\usepackage{bm,texdraw,times}
\begin{document}
\title{Comment on ``Renormalization-group picture of the Lifshitz
  critical behavior''}

\author{H.~W. Diehl}
\author{M. Shpot}
\altaffiliation{Permanent address: Institute of Condensed Matter Physics, 1 Svientsitskii str., 79011 Lviv, Ukraine}
\affiliation{%
Fachbereich Physik, Universit{\"a}t Essen, 45117 Essen,
    Federal Republic of Germany
}%

\date{\today}

\begin{abstract}
  We show that the recent renormalization-group analysis of Lifshitz
  critical behavior presented by Leite [Phys.\ Rev.~B {\bf 67}, 104415
  (2003)] suffers from a number of severe deficiencies. In particular,
  we show that his approach does not give an ultraviolet finite
  renormalized theory, is plagued by inconsistencies, misses the
  existence of a nontrivial anisotropy exponent $\theta\ne 1/2$, and
  therefore yields incorrect hyperscaling relations. His
  $\epsilon$-expansion results to order $\epsilon^2$ for the critical
  exponents of $m$-axial Lifshitz points are incorrect both in the
  anisotropic ($0<m<d$) and the isotropic cases ($m=d$). The inherent
  inconsistencies and the lack of a sound basis of the approach makes
  its results unacceptable even if they are interpreted in the sense
  of approximations.
\end{abstract}

\maketitle
Leite\cite{Lei03} recently formulated a new renormalization-group (RG)
picture of Lifshitz critical behavior. This work is built on his
previous one\cite{AL01} in which $\epsilon$-expansion results to order
$\epsilon^2$ were reported for the critical exponents $\nu_{L2}$,
$\eta_{L2}$ and $\gamma_{L}$ of an $m$-axial Lifshitz point. We have
pointed out elsewhere\cite{DS01a} that these results, which are in
conflict with ours,\cite{DS00a,SD01,DS02} are incorrect due to an
erroneous evaluation of Feynman integrals. While the main points of
our criticism in Ref.~\onlinecite{DS01a} apply equally well to
Ref.~\onlinecite{Lei03}, the latter makes mistakes on an even more
basic level, as is discussed below.

(i) Leite's renormalization scheme \emph{does not yield an ultraviolet
  (uv) finite (renormalized) theory}, and \emph{the structure of the
  RG he formulates is incorrect.}

To see this, note that in dealing with the anisotropic case ($m\ne
0,\,d$), he introduces a renormalization of the bare coupling constant
$u_0$ and two renormalization factors $Z_\phi$ and $Z_{\phi^2}$ to
renormalize the vertex functions $\Gamma^{(N,L)}$ with $N>0$ $\phi$
fields and $L$ insertions of $\phi^2$. Thus a \emph{single}
renormalization factor $Z_\phi$ is available to absorb the
$\bm{q}$-dependent primitive uv divergences of
$\Gamma^{(2,0)}(\bm{q})$, where
$\bm{q}=(\bm{k},\bm{p})\in{\mathbb{R}}^m\times{\mathbb{R}}^{d-m}$ is a
$d$-dimensional momentum. Yet, both
$\partial\Gamma^{(2,0)}(\bm{k},\bm{p})/\partial k^4$ as well as
$\partial\Gamma^{(2,0)}(\bm{k},\bm{p})/\partial p^2$ are primitively
divergent. Using his normalization conditions (2a)--(2e), one can
determine $Z_\phi$ such that the uv singularity $\sim p^2/\epsilon$ of
$\Gamma^{(2,0)}$ gets absorbed by $Z_\phi$, i.e., via the counterterm
$(Z_\phi-1)\,{\left|\nabla_{(d-m)}\phi\right|}^2/2$. But a pole $\sim
k^4/\epsilon$ will then remain in his ``renormalized''
$\Gamma^{(2,0)}_R(\bm{k},\bm{p})$ because, with this choice of
$Z_\phi$, the counterterm
$(Z_\phi-1)\,{\left|\nabla_m^2\phi\right|}^2/2$ does \emph{not} cancel
this divergence, as can be seen from our result (B4) in
Ref.~\onlinecite{SD01}
for the graph %
\raisebox{-2.0pt}{\begin{texdraw} \drawdim pt \setunitscale 2.2 \linewd
    0.3 \move(-3 0) \move(5 0) \lellip rx:5 ry:2 \move(-2 0) \rlvec(14
    0) \move(13 0)
\end{texdraw}}. %
Conversely, if $Z_\phi$ is determined so as to cancel the divergence
$\sim k^4/\epsilon$, employing, e.g., his normalization conditions
(4a)--(4c), then a term $\sim p^2/\epsilon$ will remain in
$\Gamma^{(2,0)}_R(\bm{k},\bm{p})$. That not both primitive divergences
can be absorbed by a single renormalization factor $Z_\phi$ is borne
out by the fact that the renormalization factors associated with the
above counterterms, called $Z_\phi$ and $Z_\phi Z_\sigma$ in
Ref.~\onlinecite{SD01} and explicitly given in its equations (40) and
(41), differ. [Fixing them by appropriate normalization condition
rather than by minimal subtraction of poles would change their
regular, but not their singular, parts.]

Hence Leite's renormalized function $\Gamma_R^{(2,0)}$ is ill-defined,
and since the renormalization parts of $\Gamma^{(2,0)}$ appear as
divergent subgraphs of other vertex functions, his ``renormalized''
theory quite generally has this deficiency. The fact that the
counterterms he employs are insufficient to subtract all primitive
$\bm{q}$-dependent divergences of $\Gamma^{(2,0)}(\bm{q})$ implies
that uv singular pieces of \emph{nonlocal} form produced by
higher-order graphs containing the subgraph\cite{rem:si,DRG03}
\raisebox{-2.0pt}{\begin{texdraw} \drawdim pt \setunitscale 2.2 \linewd
    0.3 \move(-3 0) \move(5 0) \lellip rx:5 ry:2 \move(-2 0) \rlvec(14
    0) \move(13 0)
\end{texdraw}}, %
such as 
\raisebox{-2.0pt}{\begin{texdraw} \drawdim pt
    \setunitscale 2.2 \linewd 0.3 \move(-3 0) \move(5 0) \lellip rx:6
    ry:2 \move(5 2) \lellip rx:4 ry:2 \move(-1 0) \rlvec(-1 1)
    \move(-1 0) \rlvec(-1 -1) \move(11 0)\rlvec(1 1) \move(11 0)
    \rlvec(1 -1) \move(13.5 0)
\end{texdraw}}, %
\emph{will not get canceled by the subtractions provided by the  counterterms to two-loop order}.

(ii) Leite's insufficient choice of counterterms \emph{is biased
  towards giving the incorrect value $\theta=1/2$ for the anisotropy
  exponent $\theta=\nu_{L4}/\nu_{L2}$.}

That is, if his renormalization scheme worked, rather than being
plagued by the unacceptable inconsistencies (i), the ratio of the
renormalization factors associated with the counterterms $\propto
{\left|\nabla_m^2\phi\right|}^2$ and $\propto
{\left|\nabla_{(d-m)}\phi\right|}^2$, i.e., the renormalization factor
$Z_\sigma$ of Ref.~\onlinecite{SD01}, would have to be uv finite. This
in turn would imply the value $\theta=1/2$ for the anisotropy exponent
to all orders in $\epsilon$. Indeed, Leite finds the value $1/2$ for
both $\nu_{l4}/\nu_{l2}=\theta$ and
$\eta_{L2}/\eta_{L4}.$\cite{rem:theta} Yet, this is wrong because
$Z_\sigma$ \emph{must have poles in $\epsilon$} as we saw above. As a
consequence, the $\epsilon^2$ term of $\theta$ is \emph{nonzero} [cf.\ 
Eq.~(84) of Ref.~\onlinecite{DS00a} and Sect.~4 of
Ref.~\onlinecite{SD01}].

(iii) Leite obtained \emph{incorrect hyperscaling relations} because
he missed the fact that $\theta$ is an independent exponent, not
identical to $1/2$ for all $\epsilon>0$.

For example, his results (54a)--(54d) for $\delta_{L2}$, $\beta_{L2}$,
$\delta_{L4}$, and $\beta_{L4}$ do not hold. These relations violate
standard scaling laws such as
\begin{equation}
  \label{eq:betal2}
\beta_{L2}=\frac{\nu_{L2}}{2}[d-2+\eta_{L2}+m\,(\theta-1)]
\end{equation}
whenever $\theta\ne 1/2$.\cite{rem:hyperscaling}

(iv) The author apparently \emph{misunderstands} the \emph{role played
  by the variable} $\sigma$, as his remarks in the last paragraph of
Sect.~III and in Sect.~VI.A indicate.

Since the classical scaling dimensions of the momentum components $p$
and $k$ differ, a dimensionful parameter $\sigma$ is indeed needed to
relate them: $\sigma^{1/2}\,k^2$ and $p$ both have the dimension
(length)$^{-1}$. It is true that $\sigma$ could be set to unity.
However, the important point the author misses is that an initial
value $\sigma=1$ gets mapped under RG transformations
$\mu\to\mu\,\ell$ onto a scale-dependent one\cite{DS00a,SD01}
$\bar{\sigma}(\ell)$ \emph{different from unity}.

(v) The author's $O(\epsilon^2)$ results for the critical exponents
of the \emph{isotropic Lifshitz point} ($d=m$) \emph{are also false};
the discrepancies with known results\cite{DS02,HLS75a} are again
\emph{due to his incorrect calculation of  Feynman integrals}.

In his treatment of these integrals, he constrains internal momenta
over which one must integrate to be orthogonal to other, external
momenta. As a consequence he gets even the simple one-loop integral
$I_2(K)$ defined in Eq.~(150) wrong. The error occurs already in the
transition to Eq.~(151). Similar ``approximations'' (mistakes) are
made in his calculation of two-loop integrals. He asserts that our
results in Ref.~\onlinecite{DS02} could not be trusted because we
absorbed a convenient factor $F_d$ in the coupling constant. He is
quite mistaken: The choice of such a factor corresponds to an uv
finite reparametrization of the theory which does not affect universal
quantities.

(vi) We \emph{fail to see} that Leite's (incorrect)
$\epsilon$-expansion results \emph{qualify as acceptable
  approximations}.

Being unaware of the fundamental problems of his approach mentioned
above, he obviously thinks that his $\epsilon$-expansion results are
correct despite the approximations he made in his computation of
Feynman integrals. Evidently, this is \emph{not} the case.

We are convinced that the property of the dimensionality expansion to
yield \emph{asymptotically exact} series expansions is an extremely
valuable one which \emph{should not be sacrificed} except for
compelling reasons.  Nevertheless one may ask whether Leite's results
(or small modifications thereof) might be acceptable when interpreted
in the sense of approximations, even though we see no need for
approximate $\epsilon$-expansion results. We believe that any such
approximation scheme ought to meet two important criteria: (a) It
should be justifiable by convincing physical and/or mathematical
reasons; (b) it should be consistent and yield a well-defined
approximate theory.

From our above critique it is clear that neither (a) nor (b) is
fulfilled by Leite's analysis. Note that the goal (b) is not at all
trivial to achieve when following the rationale of defining an
approximate renormalized theory. If one determines counterterms such
that they absorb the uv singularities of approximate Feynman integrals
of the corresponding renormalization parts --- rather than those of
the true ones ---,\cite{Zim70,ZJ96} one inevitably runs into the
problem that these renormalization parts also appear as subgraphs of
higher-order graphs of the same and other vertex functions. Since the
approximately determined lower-order counterterms do not cancel the
true uv singularities of these subgraphs, nonlocal uv singularities
generally will remain unless one succeeds in designing an
approximation scheme that produces approximate expressions for Feynman
integrals of, in principle, \emph{arbitrary order} which comply with
the local structure of their primitive uv singularities, so that a
well-defined uv finite approximate renormalized theory results.

As long as one works with the correct, unapproximated Feynman
integrals, the uv finiteness of the theory can be proven with the aid
of the forest formula\cite{Zim70} by explicitly giving the
subtractions that a general Feynman integral requires to render it uv
finite and to relate the final subtractions of the primitively
divergent graphs to the theory's counterterms. In order to be sure
that the approximation scheme yields a well-defined renormalized
theory, one would have to extend such proofs to the approximated
theory or at least present convincing evidence for its
renormalizability. Depending on the choice of approximation scheme, a
mathematically rigorous proof may well turn out to be more involved
than familiar renormalizability proofs of the proper, unapproximated
theory.

In summary, Leite's analysis has no sound basis, is plagued by
inconsistencies and uv problems, and his results are incorrect,
failing even to qualify as exceptable approximations.

\begin{acknowledgments}
  We gratefully acknowledge support by the Deutsche
  Forschungsgemeinschaft (DFG) via grant \# Di-378/3.
\end{acknowledgments}
%
%\bibliography{bank,remdscom}

\begin{thebibliography}{13}
\bibitem[{rem({\natexlab{a}})}]{rem:si}
\bibinfo{note}{The bulk counterterm $\propto (Z_\phi Z_\sigma-1)$ is also
  needed to cancel the nonlocal singularities one encounters at two-loop order
  in semi-finite extensions of the model considered by Leite; see
  Ref.~\onlinecite{DRG03} for details.}
\bibitem[{rem({\natexlab{b}})}]{rem:theta}
\bibinfo{note}{In general, one has $\frac{2-\eta_{L2}}{4-\eta_{L4}}=\theta$. If
  $\theta$ were $1/2$, this would imply $\eta_{L2}/\eta_{L4}=1/2$.}
\bibitem[{rem({\natexlab{c}})}]{rem:hyperscaling}
\bibinfo{note}{The correct hyperscaling relations [e.g., Eq.~(\ref{eq:betal2})]
  can, of course, easily be derived also by phenomenological scaling
  arguments.}
\end{thebibliography}
%

\end{document}